*Commentary*

# Digital Hydrogen Platform (*DigHyd*): A Rigorously Curated Database for Hydrogen Storage Materials Empowered by AI-Assisted Literature Mining

*Seong-Hoon Jang[1], Di Zhang[*,1,2], Xue Jia[1], Hung Ba Tran[1], Linda Zhang[1,2], Ryuhei Sato[3], Yusuke Hashimoto[2], Toyoto Sato[4], Kiyoe Konno[5], Shin-ichi Orimo[*,1,4], and Hao Li[*,1]*

[1] Advanced Institute for Materials Research (WPI-AIMR), Tohoku University, Sendai 980-8577, Japan

[2] Frontier Research Institute for Interdisciplinary Sciences (FRIS), Tohoku University, Sendai 980-8577, Japan

[3] Department of Materials Engineering, The University of Tokyo, Tokyo 113-8656, Japan

[4] Institute for Materials Research, Tohoku University, Sendai, 980-8577

[5] Institute of Fluid Science, Tohoku University, Sendai, 980-8577, Japan

*Corresponding authors:

di.zhang.a8@tohoku.ac.jp (D. Zhang)

shin-ichi.orimo.a6@tohoku.ac.jp (S. Orimo),

li.hao.b8@tohoku.ac.jp (H. Li)

ABSTRACT

Solid-state hydrogen storage materials are promising candidates for safe and compact hydrogen storage; however, data-driven discovery in this field remains limited by the availability of large-scale, well-curated datasets. Here, we present the Digital Hydrogen Platform (*DigHyd*: www.dighyd.org), a rigorously curated database comprising > 4,000 experimental literature sources and >30,000 data entries on hydrogen storage materials, constructed through AI-assisted literature mining combined with human-in-the-loop validation. In addition to gravimetric hydrogen storage density ($w$), *DigHyd* also covers thermodynamic parameters, specifically the enthalpy ($\Delta H$) and entropy ($\Delta S$) changes associated with hydrogenation reactions, primarily defined as $M + \frac{1}{2}\text{H}_2 \rightleftharpoons M\text{H}$. These parameters were obtained by manually analyzing multi-temperature pressure-composition-temperature (PCT) data using van't Hoff analysis. By focusing on $\Delta H$ and $\Delta S$ rather than fixing equilibrium pressure at a single temperature, *DigHyd* enables flexible evaluation of equilibrium behavior under application-specific operating conditions. Statistical analyses reveal distinct distributions of thermodynamic parameters across material classes, together with broad compositional variability within representative hydride systems. Furthermore, both physically interpretable symbolic regression and black-box XGBoost models achieve comparable predictive performance for $w$ and equilibrium pressure at room temperature ($P_{\text{eq,RT}}$), demonstrating internal consistency and learnable composition–property relationships within the curated dataset. Overall, *DigHyd* provides a rigorously curated thermodynamic dataset that serves as a reliable basis for data-driven analyses of hydrogen storage materials and supports systematic exploration of structure–property relationships.

# 1. Introduction

Hydrogen is widely recognized as a key energy carrier for a future low-carbon society due to its high gravimetric energy density and clean combustion.[1-4] A central challenge for hydrogen-based energy systems, however, lies in the development of safe, compact, and reversible storage technologies. Among various approaches, solid-state hydrogen storage using hydride materials has attracted sustained interest because of its inherent safety, high volumetric density, and potential for reversible operation under moderate conditions.[5-8]

Hydride materials encompass a broad range of material classes, including interstitial metal hydrides, saline (ionic) hydrides, complex hydrides, multi-component or destabilized systems, and porous materials such as metal-organic frameworks. This diversity provides a rich materials space but also complicates the establishment of general design principles. In contrast to compressed or liquefied hydrogen storage, the performance of hydride-based systems is governed not only by storage capacity but also by thermodynamic properties that dictate equilibrium pressure, operating temperature, and reversibility.

Despite decades of experimental research, existing hydrogen storage datasets suffer from two persistent limitations. First, gravimetric hydrogen densities ($w$) are often reported inconsistently, with insufficient distinction between theoretical capacities calculated directly from the stoichiometric chemical formula, reversible capacities measured during hydrogen absorption-desorption cycling, and experimentally accessible capacities obtained under specific pressure and temperature conditions.[9,10] Second, the thermodynamic quantities that fundamentally control equilibrium behavior, namely the enthalpy ($\Delta H$) and entropy changes ($\Delta S$) of hydrogenation reactions, are not systematically reported across the literature at the scale required for large databases, limiting their direct use in data-driven materials analysis.

Recent advances in artificial intelligence (AI), particularly large language models (LLMs), have enabled automated extraction of materials data from the literature at unprecedented scale.[11-15] However, while AI-based approaches excel at collecting large volumes of reported values, they are insufficient on their own for ensuring numerical correctness. First, special care is required when curating $w$, as some literature reports hydrogen content in terms of the number of hydrogen atoms per metal atom rather than gravimetric hydrogen storage density. Second, the extraction and interpretation of thermodynamic parameters, namely $\Delta H$ and $\Delta S$, from pressure-composition-temperature (PCT) measurements often require expert judgment, particularly when based on multi-temperature datasets, as well as careful unit unification to ensure consistency across data points.

To address these issues, in this work, we introduce the **Digital Hydrogen Platform (*DigHyd*:** [www.dighyd.org](www.dighyd.org)**)**, a rigorously curated database of metal hydrides constructed using an AI-assisted, human-in-the-loop framework.[14, 16, 17] The central design philosophy of *DigHyd* is to consistently curate not only $w$ but also the thermodynamics parameters $\Delta H$ and $\Delta S$ changes of hydrogenation reactions rather than storing equilibrium pressures at a specific temperature. This design enables flexible evaluation of equilibrium behavior under diverse operating conditions and provides a physically grounded foundation for data-driven modeling. Through statistical analysis and comparative machine learning studies, we demonstrate that *DigHyd* captures both the diversity and internal coherence necessary for systematic exploration of composition-property relationships in hydrogen storage materials.

## 2. Methodology

The construction of the *DigHyd* platform began with large-scale literature mining using an AI-assisted workflow.[14, 16, 17] Scientific articles related to hydrogen storage and metal hydrides were identified through keyword-based searches and bibliographic databases. LLM-based tools were employed to extract candidate information from published articles, including chemical composition, reported gravimetric hydrogen storage density, PCT data, or thermodynamic quantities where available. These AI-extracted values were treated as preliminary candidates and subjected to further validation rather than being directly incorporated into the database, while retaining the associated DOI to ensure traceability.

The core of *DigHyd* lies in the rigorous manual curation of thermodynamic parameters governing hydrogen absorption and desorption, following AI-assisted data mining: $w$, $\Delta H$, and $\Delta S$. For instance, thermodynamic quantities $\Delta H$ and $\Delta S$ associated with metal hydrides (while *DigHyd* is not limited to this materials space) are defined with respect to the hydrogenation reaction:

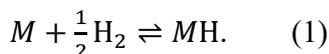

$$M + \frac{1}{2}H_2 \rightleftharpoons MH. \quad (1)$$

For materials with reported multi-temperature PCT data, equilibrium plateau pressures were identified at each temperature. Van't Hoff analysis was then performed according to:

$$\ln P = -\frac{\Delta H}{RT} + \frac{\Delta S}{R}, \quad (2)$$

where $P$ is the equilibrium pressure, $T$ is the absolute temperature, and $R$ is the gas constant. $\Delta H$ was extracted from the slope of the linear fit, and $\Delta S$ from the intercept. In cases where $\Delta H$ and $\Delta S$ were directly reported, the values were manually cross-checked against the corresponding PCT data when available, as presented in the figures. This human-in-the-loop validation step was essential to ensure thermodynamic reliability. Overall, for metal hydride cases only, 4,643 data points of $w$ and 652 data pairs of $\Delta H$ and $\Delta S$ were obtained through subsequent manual curation.

## 3. Results and Discussion

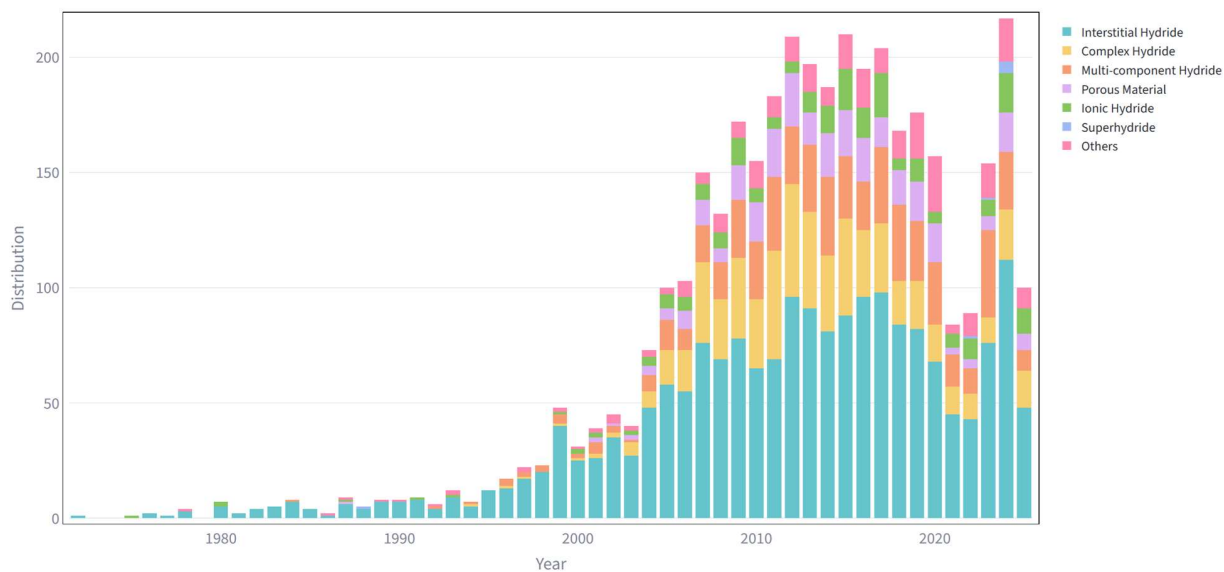

**Fig. 1.** Temporal distribution of hydrogen storage materials in the *DigHyd* platform classified by material type. The stacked bar chart shows the number of reported materials per year, grouped into interstitial hydrides, complex hydrides, multi-component hydrides, porous materials, ionic hydrides, superhydrides, and others, illustrating the evolution of research focus across different classes over time. Reproduced from Ref. 14, under the terms of the Creative Commons CC BY-NC license.

**Fig. 1** presents the temporal distribution of hydrogen storage materials in the *DigHyd* platform, categorized by material class. A steady increase in the number of reported materials is observed from the early stages of hydrogen storage research to the mid-2010s, reflecting sustained and growing interest in solid-state hydrogen storage. Interstitial hydrides constitute the largest fraction throughout most of the timeline, consistent with their long-standing role as model systems for reversible hydrogen absorption and desorption. From the late 1990s onward, the emergence and subsequent growth of complex hydrides, multi-component systems, and porous materials indicate a diversification of research directions aimed at improving storage capacity and tunability. More recent years show continued contributions across multiple material classes, highlighting the broadening scope of hydrogen storage research rather than a shift toward a single dominant material type. Overall, this figure demonstrates that the *DigHyd* database captures both the historical development and the evolving diversity of hydrogen storage materials reported in the literature.

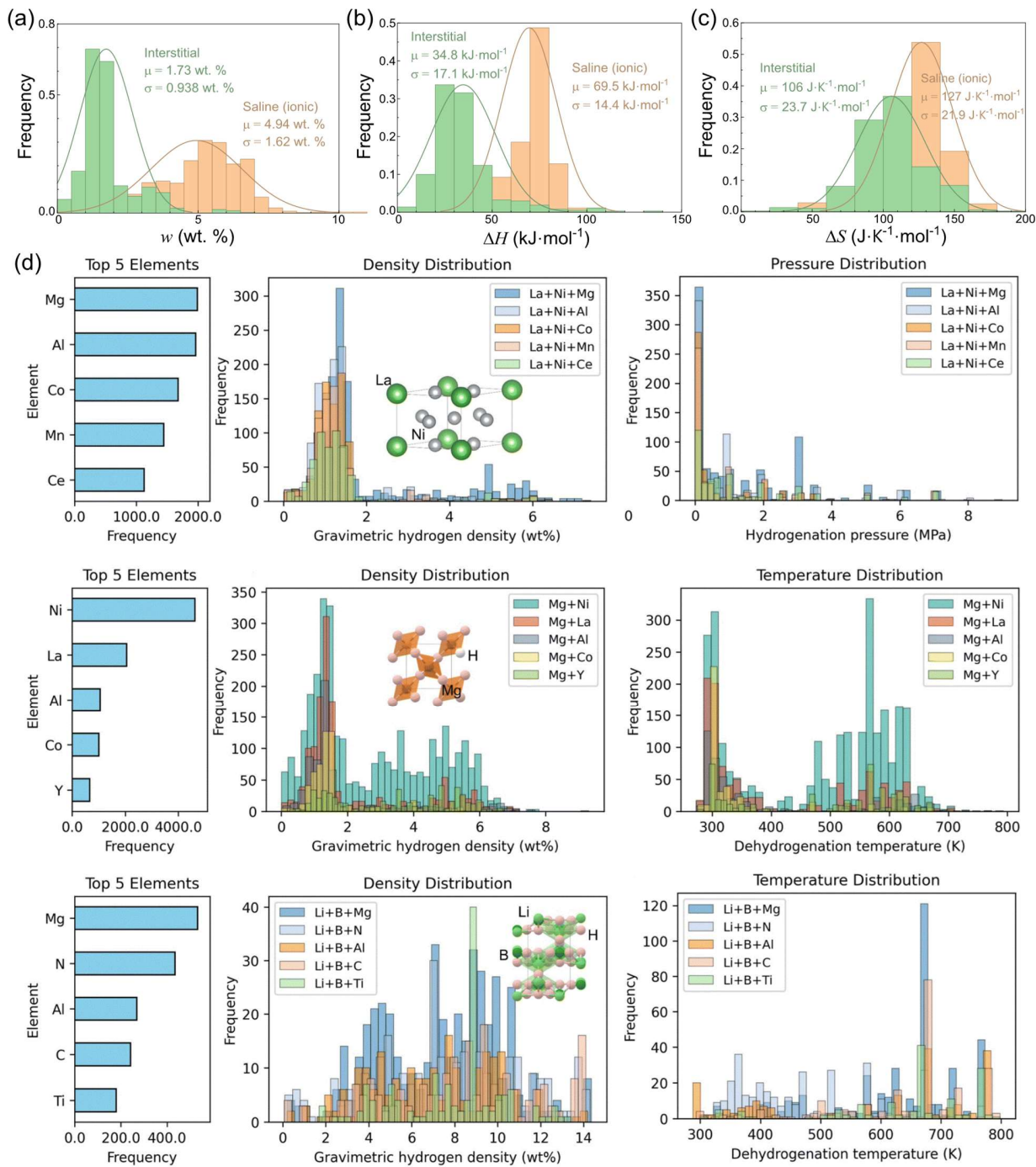

**Fig. 2.** (a-c) Distributions of key hydrogen storage properties for interstitial (represented by green blocks) and saline (ionic; by yellow blocks) metal hydrides in the *DigHyd* database. From top to bottom, the panels show histograms of (a) gravimetric hydrogen storage density ($w$), (b) enthalpy change ($\Delta H$), and (c) entropy change ($\Delta S$) associated with hydrogenation reactions. Solid curves represent Gaussian fits to the distributions, with the mean ($\mu$) and standard deviation ($\sigma$) indicated for each material class. (d) Representative compositional modification trends in $LaNi_5$-, $MgH_2$-, and $LiBH_4$-based systems, showing frequently added elements and their influence on hydrogen storage density, equilibrium pressure, and (de)hydrogenation temperature. Reproduced from Ref. 14 under the terms of the Creative Commons CC BY-NC license.

**Figs. 2a-c** compares the statistical distributions of $w$, $\Delta H$, and $\Delta S$ for interstitial and saline (ionic) metal hydrides in the *DigHyd* database. For $w$, 2,859 data points for interstitial cases and 1,785 data points for saline (ionic) cases were included; for $\Delta H$ and $\Delta S$, 371 data points for interstitial cases and 281 data points for saline (ionic) cases were included. Interstitial hydrides exhibit relatively low $w$ with a narrow distribution centered at approximately 1.73 wt.% (**Fig. 2a**). In contrast, saline hydrides show substantially higher $w$ and a broader distribution, consistent with their stronger ionic metal-hydrogen bonding. The $\Delta H$ distributions further distinguish the two classes: interstitial hydrides cluster around moderate $\Delta H$ values, whereas saline hydrides are shifted toward significantly higher $\Delta H$, indicating more energetically demanding hydrogen release (**Fig. 2b**). Despite this clear separation in $\Delta H$, the $\Delta S$ distributions partially overlap, indicating that entropy changes are more strongly influenced by the contribution of gaseous $H_2$ rather than by the specific bonding nature of the hydride (**Fig. 2c**).

It is noteworthy that materials with comparable $\Delta S$ can exhibit substantially different equilibrium pressures at a given temperature due to differences in $\Delta H$. As a result, such materials may display distinct hydrogen storage behavior within a target operating temperature and pressure window, which the *DigHyd* database enables users to assess in a flexible manner. This observation underscores the necessity of simultaneously considering $\Delta H$ and $\Delta S$ when evaluating hydrogen storage performance, particularly for applications near ambient conditions, in addition to $w$.

In addition to class-level thermodynamic distinctions, the *DigHyd* database also captures substantial compositional diversity within representative material families, as illustrated in **Fig. 2d**. By analyzing frequently modified base systems such as $LaNi_5$, $MgH_2$, and $LiBH_4$ as representative examples, the database reveals the wide range of elemental substitutions and additives explored in the literature. Rather than focusing on isolated case studies, this analysis statistically aggregates the most commonly introduced elements and maps their associated distributions of gravimetric hydrogen storage density, equilibrium pressure, and (de)hydrogenation temperature. The broad spread of these distributions demonstrates that even within a single prototype material, hydrogen storage behavior spans a wide thermodynamic and kinetic space depending on compositional tuning.

This compositional analysis highlights the information diversity embedded in *DigHyd*. The database does not merely catalog distinct material classes (*e.g.*, interstitial *vs.* ionic hydrides), but also records systematic modification strategies within each class, reflecting decades of experimental optimization efforts. Such diversity enables users to explore composition–property relationships at multiple levels: from class-wide thermodynamic trends (**Fig. 2a-c**) to fine-grained alloying effects within canonical systems (**Fig. 2d**). Consequently, *DigHyd* provides a multidimensional dataset that captures both inter-class contrasts and intra-class variability, forming a robust basis for data-driven hydrogen storage materials design.

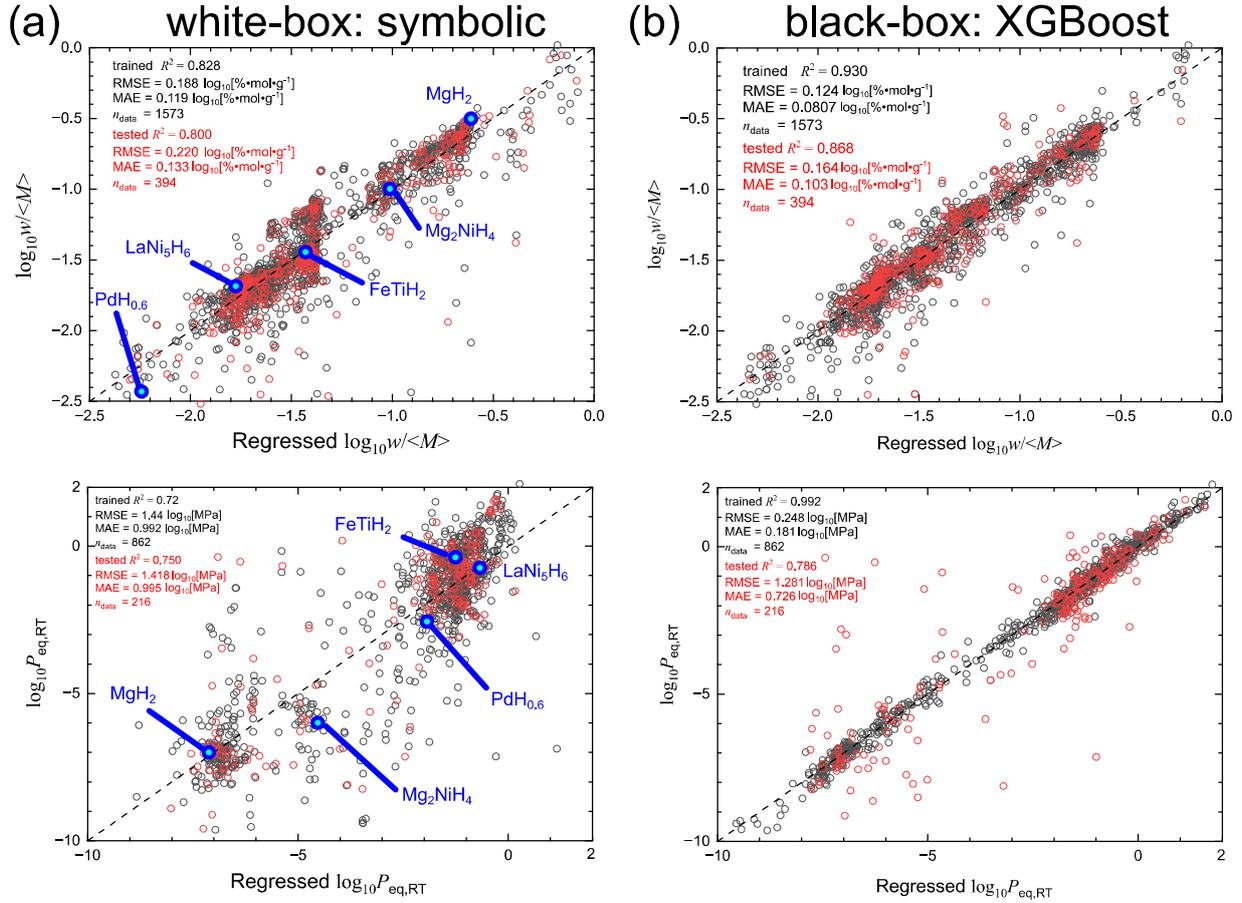

**Fig. 3.** Comparison between physically interpretable and black-box machine learning models for hydrogen storage properties. (a) White-box symbolic regression models and (b) XGBoost models trained on the curated *DigHyd* dataset. In each panel, the upper inset shows predicted *versus* true gravimetric hydrogen storage density ($w$), and the lower inset shows predicted *versus* true equilibrium pressure at room temperature ($P_{eq,RT}$). Training and test datasets follow an 80:20 split and are denoted by black and red open circles, respectively. Five representative hydrides ($MgH_2$, $Mg_2NiH_4$, $FeTiH_2$, $PdH_{0.6}$, and $LaNi_5H_6$) are highlighted for reference. $R^2$, RMSE, and MAE values for the train and test datasets of each case are shown. Reproduced in part from Ref. 16 under the terms of the Creative Commons CC BY-NC license, with additional XGBoost results included.

For subsequent regression modeling, we introduce the equilibrium pressure at room temperature ($P_{eq,RT}$) as a derived thermodynamic descriptor calculated from the curated enthalpy ($\Delta H$) and entropy ($\Delta S$) values *via* the van't Hoff relation, as expressed in **Eq. (2)**. Instead of modeling $\Delta H$ and $\Delta S$ separately, $P_{eq,RT}$ serves as a practically meaningful target because it directly reflects the operating pressure required under near-ambient conditions. From an application perspective, equilibrium pressure at a specified temperature integrates both enthalpic and entropic contributions into a single descriptor that determines whether a material operates within a desirable pressure window. For this reason, $P_{eq,RT}$ was selected as a regression target alongside $w$, enabling property prediction in a form directly relevant to engineering considerations.

The predictive performances of the physically interpretable white-box symbolic regression model (**Fig. 3a**)[16, 18] and the black-box XGBoost model (**Fig. 3b**)[19] are comparable, with similar $R^2$, RMSE, and MAE values for the test datasets of both $w$ and $P_{eq,RT}$. This consistency across modeling paradigms indicates that the curated dataset contains learnable and internally coherent composition-property relationships. While XGBoost offers flexible nonlinear fitting, the symbolic regression approach provides explicit analytical expressions with physically meaningful descriptors, allowing direct interpretation of how elemental features influence hydrogen storage behavior. The comparable accuracy combined with enhanced interpretability highlights the advantage of coupling thermodynamically rigorous data curation with physically transparent modeling frameworks, and further supports the structural consistency of the *DigHyd* dataset.

## 3. Conclusion and Perspectives

In summary, the *DigHyd* platform provides a systematically curated and thermodynamically grounded resource for hydrogen storage materials research. By capturing the temporal evolution and material diversity of the literature, *DigHyd* reflects both the historical focus on interstitial metal hydrides and the gradual expansion toward more complex and diverse material classes. The comparative analysis of gravimetric hydrogen storage density and thermodynamic parameters highlights systematic differences between hydride classes. In addition, compositional mapping within representative systems such as $LaNi_5$-, $MgH_2$-, and $LiBH_4$-based materials highlights substantial intra-class variability arising from systematic alloying and additive strategies.

Furthermore, the ability of composition-based machine learning models to reproduce $w$ and $P_{eq,RT}$ trends supports the internal coherence of the curated dataset and illustrates its suitability for data-driven analysis. The comparable predictive performance of physically interpretable symbolic regression and black-box XGBoost models further demonstrates that the dataset contains learnable and physically meaningful composition-property relationships. Looking forward, the *DigHyd* framework provides a foundation for more refined thermodynamic screening under application-specific operating conditions and offers a scalable pathway toward integrating additional materials classes and properties as high-quality experimental data become available.

While the current thermodynamic curation primarily focuses on metal hydrides, the framework is designed to be extensible to other hydrogen storage material classes as high-quality data become available. Hence, future extensions of *DigHyd* will include the rigorous curation of porous and complex hydrides, incorporation of kinetic descriptors, and continued development toward physically constrained AI Agents for hydrogen storage materials discovery. By emphasizing thermodynamic correctness over superficial simplicity, *DigHyd* aims to support more reliable and insightful data-driven research in hydrogen storage science.

**Associated content**

**Author information**

**Data Availability**

**Acknowledgments**

This work was supported by The Green Technologies of Excellence (GteX) Program Japan Grant No. JPMJGX23H1.

**TOC**

Literature: hydrogen storage materials

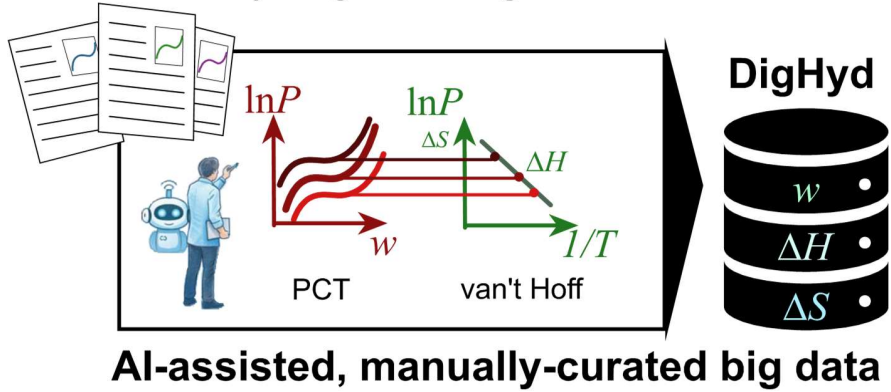

**AI-assisted, manually-curated big data**